\documentclass[aps,pra,reprint,showpacs,floatfix,a4paper,superscriptaddress]{revtex4-1}
\usepackage{graphicx}
\usepackage{amsmath}
\usepackage{amssymb}
\usepackage{amsfonts}
\usepackage{color}
\usepackage{relsize}
\usepackage{txfonts}
\usepackage[T1]{fontenc}
\usepackage{microtype}

\renewcommand\Re{\operatorname{Re}}

\newcommand\maybeflushthispage{\vfil\penalty1000\vfilneg}

\begin{document}

\title{Optomechanically induced transparency in membrane-in-the-middle setup at room temperature}

\author{M. Karuza}
\affiliation{School of Science and Technology, Physics Division, University of Camerino, via Madonna delle Carceri, 9, I-62032 Camerino (MC), Italy, and INFN, Sezione di Perugia, Italy}
\affiliation{Department of Physics and Center for Micro and Nano Sciences and Technologies, Radmile Matejcic 2, HR-51000 Rijeka, Croatia}
\author{C. Biancofiore}
\affiliation{School of Science and Technology, Physics Division, University of Camerino, via Madonna delle Carceri, 9, I-62032 Camerino (MC), Italy, and INFN, Sezione di Perugia, Italy}
\author{M. Bawaj}
\affiliation{School of Science and Technology, Physics Division, University of Camerino, via Madonna delle Carceri, 9, I-62032 Camerino (MC), Italy, and INFN, Sezione di Perugia, Italy}
\author{C. Molinelli}
\affiliation{School of Science and Technology, Physics Division, University of Camerino, via Madonna delle Carceri, 9, I-62032 Camerino (MC), Italy, and INFN, Sezione di Perugia, Italy}
\author{M. Galassi}
\affiliation{School of Science and Technology, Physics Division, University of Camerino, via Madonna delle Carceri, 9, I-62032 Camerino (MC), Italy, and INFN, Sezione di Perugia, Italy}
\author{R. Natali}
\affiliation{School of Science and Technology, Physics Division, University of Camerino, via Madonna delle Carceri, 9, I-62032 Camerino (MC), Italy, and INFN, Sezione di Perugia, Italy}
\author{P. Tombesi}
\affiliation{School of Science and Technology, Physics Division, University of Camerino, via Madonna delle Carceri, 9, I-62032 Camerino (MC), Italy, and INFN, Sezione di Perugia, Italy}
\author{G. Di Giuseppe}
\affiliation{School of Science and Technology, Physics Division, University of Camerino, via Madonna delle Carceri, 9, I-62032 Camerino (MC), Italy, and INFN, Sezione di Perugia, Italy}
\author{D. Vitali}
\affiliation{School of Science and Technology, Physics Division, University of Camerino, via Madonna delle Carceri, 9, I-62032 Camerino (MC), Italy, and INFN, Sezione di Perugia, Italy}

\begin{abstract}
We demonstrate the analogue of electromagnetically induced transparency in a room temperature cavity optomechanics setup formed by a thin semitransparent membrane within a Fabry-P\'{e}rot cavity. Due to destructive interference, a weak probe field is completely reflected by the cavity when the pump beam is resonant with the motional red sideband of the cavity. Under this condition we infer a significant slowing down of light of hundreds of microseconds, which is easily tuned by shifting the membrane along the cavity axis. We also observe the associated phenomenon of electromagnetically induced amplification which occurs due to constructive interference when the pump is resonant with the blue sideband.
\end{abstract}

\pacs{42.50.Gy, 42.50.Wk, 85.85.+j,42.50.Ex}

\maketitle

Cavity optomechanics is currently a very active field of investigation owing to the disparate possibilities offered by the ability to manipulate the state and dynamics of nanomechanical resonator with light, and at the same time of controlling light by tailoring its interaction with one (or more) mechanical resonances~\cite{Kippenberg2007,Genes2009,Marquardt2009,Favero2009,Aspelmeyer2012}. A notable example of this kind of light beam control is provided by the optomechanical analogue of electromagnetically induced transparency (EIT)~\cite{Arimondo1996,Fleischhauer2005}, the so called optomechanically induced transparency (OMIT), which has been recently demonstrated both in optical~\cite{Weis2010,Safavi-Naeini2011} and microwave domains~\cite{Teufel2011,Massel2012}. In EIT an intense control field (pump) modifies the optical response of an opaque medium making it transparent in a narrow bandwidth; the concomitant steep variation of the refractive index induces a significant slowing down of the group velocity of a probe beam~\cite{Hau1999}, which can be used to delay, stop, store and retrieve both classical~\cite{Phillips2001,Liu2001} and quantum information~\cite{Fleischhauer2005} encoded in a light field. In OMIT, the internal resonance of the medium is replaced by a dipole-like interaction of optical and mechanical degrees of freedom which occurs when the pump is tuned to the lower motional sideband of the cavity resonance. EIT has been first observed in atomic gases and more recently in a variety of solid-state systems such as quantum wells, dots and nitrogen--vacancy centers~\cite{Phillips2003,Santori2006,Xu2008}. OMIT may offer various advantages with respect to these latter implementations: i) it does not rely on naturally occurring resonances and could therefore be applied to previously inaccessible wavelength regions; ii) a single optomechanical element can already achieve unity contrast, which in the atomic case is only possible within the setting of cavity quantum electrodynamics~\cite{Mucke2010}; iii) one can achieve significant optical delay times, since they are limited only by the mechanical resonance lifetime of the optomechanical system.

With the exception of some results shown in Ref.~\cite{Safavi-Naeini2011}, previous OMIT demonstrations have been carried out in a cryogenic setup; here we show OMIT and also the associated phenomenon of electromagnetically induced amplification~\cite{Massel2011} in a room temperature optomechanical setup consisting of a thin semitransparent membrane within a high-finesse optical Fabry-P\'{e}rot (FP) cavity~\cite{Thompson2008,Wilson2009}. Our setup involves free space optics rather than guided optics as in Refs.~\cite{Weis2010,Safavi-Naeini2011}, and it operates at lower frequencies (hundreds of kHz), with respect to the MHz-GHz regime of Refs.~\cite{Weis2010,Safavi-Naeini2011}, allowing us to attain significantly longer delay times, up to $1$ ms. Moreover, in Refs.~\cite{Weis2010,Safavi-Naeini2011}, the optical and mechanical modes are localized within the same structure, while in the present setup the mechanical element is separated and independent from the cavity mode, enabling the study of a larger variety of optomechanical configurations with micromechanical resonators with different material and structural properties.
While in the previous demonstrations of OMIT~\cite{Weis2010,Safavi-Naeini2011} the interference between a probe beam and a strong pump beam results in a ``transparency'' frequency window, \emph{i.e.}\ the probe beam is transmitted through the tapered optical fiber coupled to the resonator, in our system it leads to an ``opacity'' frequency window, \emph{i.e.}\ the probe is completely reflected by the cavity even if in resonance with it.
Furthermore the OMIT transparency window and the optical delay can be tuned in a simple way by properly shifting the membrane along the cavity axis.

\emph{The experimental setup}. The optical power of a laser beam at $\lambda=1\,064\,\mathrm{nm}$ produced by a Nd:YAG laser (Innolight) was distributed between a probe ($\omega_\mathrm{p}$) and a pump beam ($\omega_{\mathsmaller{\mathrm{L}}}$) by means of a cascade of a half wave plate (HWP$_1$) and a polarizing beam splitter (PBS$_1$), as shown in the setup in Fig.~1 (see also Ref.~\cite{Karuza2012}).

\begin{figure}[h]
\centering
\includegraphics[width=0.45\textwidth]{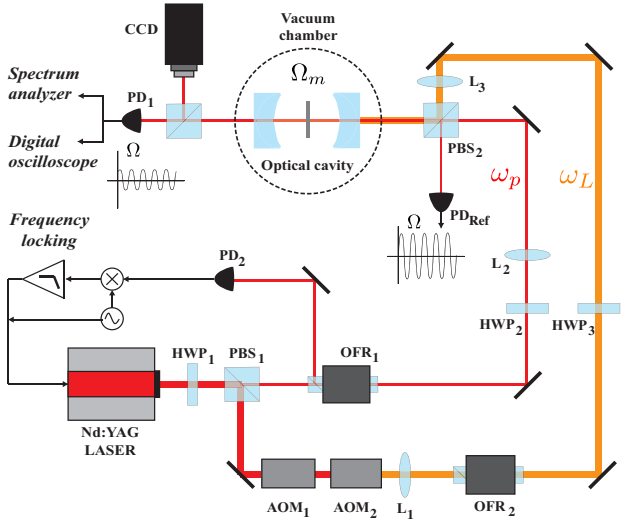}
\caption{Schematic description of the experimental setup.
 }
\label{fig:1}
\end{figure}

The probe beam power was $100\,\mu\mathrm{W}$, while the rest, that was about $200\,\mathrm{mW}$, was fed into the pump beam optical line: at the end only a small fraction of it was used.
Two cascaded acousto-optical modulators (AOMs) were used to obtain controlled frequency detuning from the probe beam in the range $0$ to $40\,\mathrm{Mhz}$, although only detunings up to $500\,\mathrm{kHz}$ have been used. The pump beam intensity is controlled by the modulation amplitude of the electrical signal used to drive AOM$_2$. After the AOMs and an optical isolator (OFR$_2$) the pump beam was mode-matched to the FP optical cavity by means of two lenses (L$_1$ and L$_3$). Before being injected in the FP cavity the pump beam was combined with the probe beam by polarization multiplexing of the fields on PBS$_2$.
The cavity was $L \approx 93\,\mathrm{mm}$ long and consisted of two equal dielectric mirrors, each with a radius of curvature $R = 10\,\mathrm{cm}$. The measured value of the empty cavity finesse was $\mathcal{F} \approx 60\,000$, consistent with the mirror's nominal reflectivity. Halfway between the mirrors a thin stoichiometric silicon nitride membrane was mounted on series of piezo-motor driven optical mounts that control the angular alignment as well as the linear positioning with respect to the optical axis.
The membrane was a commercial $1\,\mathrm{mm}\times 1\,\mathrm{mm}$ ${\rm Si_3 N_4}$ stoichiometric x-ray window (Norcada), with nominal thickness $L_{\mathrm{d}} =50\,\mathrm{nm}$, and index of refraction $n_{\mathsmaller{\mathrm{R}}} \approx 2$, supported on a $200\,\mu\mathrm{m}$ $\mathrm{Si}$ frame. It has been chosen due to its high mechanical quality factor and very low optical absorption at $\lambda = 1\,064\,\mathrm{nm}$~\cite{Zwickl2008}. Its optical properties were also experimentally verified, yielding an intensity reflection coefficient $\mathcal{R} \approx 0.18$, and an imaginary part of the index of refraction $n_{\mathsmaller{\mathrm{I}}} \approx 2 \times 10^{-6}$.
%
In order to avoid the deterioration of the mechanical properties of the membrane and optical properties of the FP cavity, the cavity was mounted inside a vacuum chamber which was evacuated by a turbo-molecular pump down to $10^{-5}\,\mathrm{mbar}$.
The probe beam light reflected from the cavity was observed by a photodiode (PD$_2$) whose output signal is amplified and fed into a frequency locking loop (described in Ref.~\cite{Karuza2012}) and a spectrum analyzer where the membrane's mechanical motion was monitored.

\emph{Langevin equation description}. The pump drives the system in a steady state characterized by the driven TEM$_{00}$ mode (with photon annihilation operator $\hat{a}$) in an intense coherent state with amplitude $\alpha_{\mathrm{s}}$, and the membrane deformed by radiation pressure. We choose the detection bandwidth in order to observe the fundamental membrane vibrational mode with resonance frequency $\Omega_{\mathrm{m}}/2\pi \approx 355.6$ kHz and quality factor $Q=\omega_{\mathrm{m}}/\gamma_{\mathrm{m}}\approx122\,000$, which we describe as a harmonic oscillator with effective mass $m$, and with dimensionless position $\hat{q}$ and momentum $\hat{p}$ satisfying the commutation rule $\left[\hat{q},\hat{p}\right]=\mathrm{i}$~\cite{Biancofiore2011,Karuza2012}. Under these conditions, dynamical effects are associated with the fluctuations $\bigl(\delta \hat{q},\delta \hat{p}\bigr)$ of the vibrational mode around its steady state $\bigl(q_{\mathrm{s}},p_{\mathrm{s}}=0\bigr)$, and with the cavity mode fluctuations $ \delta \hat{a}$ around $\alpha_{\mathrm{s}}$. This dynamics are well described by the following linearized Langevin equations~\cite{Biancofiore2011,Karuza2012}
\begin{subequations}
\label{lle}
\begin{eqnarray}
\delta \dot{\hat{q}}& =&\Omega _{\mathrm{m}}\delta \hat{p}, \label{lle1}\\
\delta \dot{\hat{p}}& =&-\left[\Omega _{\mathrm{m}}+\partial_q^2 \omega(q_{\mathrm{s}})|\alpha_{\mathrm{s}}|^2\right]\delta \hat{q}-\gamma _{\mathrm{m}}\delta \hat{p} \nonumber \\
&&-\partial_q \omega(q_{\mathrm{s}})\alpha_{\mathrm{s}} \left(\delta \hat{a} +\delta \hat{a}^{\dagger}\right) +\hat{\xi}, \label{lle2}\\
\delta \dot{\hat{a}}& =&-\left(\kappa_0+\kappa_2+\mathrm{i} \Delta \right) \delta \hat{a}-\mathrm{i} \partial_q \omega(q_{\mathrm{s}})\alpha_{\mathrm{s}} \delta \hat{q} \label{lle3} \nonumber \\
&& +
\sqrt{2\kappa_0}\hat{a}_0^{\mathrm{in}}+\sqrt{2\kappa_2}\hat{a}_2^{\mathrm{in}}+\sqrt{2\kappa_0} s_{\mathrm{p}} \mathrm{e}^{-\mathrm{i}\Omega t},
\end{eqnarray}
\end{subequations}
where we have adopted a frame rotating at the pump frequency $\omega_{\mathsmaller{\mathrm{L}}}$, and we have chosen the phase reference of the cavity field so that $\alpha _{\mathrm{s}}$ is real and positive. $\kappa_0 $ and $\kappa_2 $ denote the cavity decay rates through the input and back mirror respectively, $\hat{a}_0^{\mathrm{in}}$ and $\hat{a}_2^{\mathrm{in}}$ are the corresponding vacuum optical input white noises~\cite{Gardiner2000}, $\Delta = \omega(q_{\mathrm{s}})-\omega_{\mathsmaller{\mathrm{L}}}$ is the cavity detuning, and $\hat{\xi}$ is the thermal stochastic force. Optomechanical coupling is provided through the position-dependent cavity mode frequency, $\omega(\hat{q})=\omega_0+\Re\bigl\{\delta\omega\left[z_0(\hat{q})\right]\bigr\}$, where $\omega_0$ is the frequency in the absence of the membrane, and $\Re\bigl\{\delta\omega\left[z_0(\hat{q})\right]\bigr\}$ is the frequency shift caused by the insertion of the membrane. This shift depends on the membrane position along the cavity axis $z_0(\hat{q}) = z_0+x_0 \Theta \hat{q}$, where $z_0$ is the membrane center-of-mass position along the cavity axis, $\Theta$ is the transverse overlap integral between the optical mode and the vibrational mode~\cite{Biancofiore2011}, and $x_0 = \sqrt{\hbar/m \Omega_\mathrm{m}}$. Radiation pressure coupling is described by the first order derivative term $\partial_q \omega(q_{\mathrm{s}})$, but, as shown in Ref.~\cite{Karuza2012}, also the second-order term $\partial_q^2 \omega(q_{\mathrm{s}})$ has to be included in Eq.~(\ref{lle2}) since it accounts for an observable mechanical frequency shift which is typical for the membrane-in-the-middle setup and usually negligible in other cavity optomechanical devices.

\emph{Optomechanically induced transparency}. The last term of Eq.~(\ref{lle3}) describes the additional weak probe field of frequency $\omega_{\mathrm{p}} = \omega_{\mathsmaller{\mathrm{L}}}+\Omega$ and amplitude $s_{\mathrm{p}}$ which, together with the intense pump, induces a modulation at frequency $\Omega$ of the radiation pressure force acting on the membrane. When this modulation is close to the mechanical resonance frequency $\Omega_{\mathrm{m}}$, the vibrational mode is excited, giving rise to Stokes- and anti-Stokes scattering of light from the strong pump field. If the latter is tuned to the red sideband of the cavity, Stokes scattering is suppressed and only the anti-Stokes field at $ \omega_{\mathsmaller{\mathrm{L}}}+\Omega_{\mathrm{m}}$ builds up within the cavity. However when $\Omega \approx \Omega_{\mathrm{m}} \approx \Delta$, also the probe beam is resonant with the cavity, but destructive interference with the anti-Stokes field suppresses its build-up and as a result the probe beam is \emph{perfectly reflected} by the coupled cavity-membrane system~\cite{Weis2010,Agarwal2010}. This OMIT phenomenon is well described by the classical limit of Eqs.~(\ref{lle}), in which the fluctuation operators are replaced by classical variables. The probe modulates in time the coupled optomechanical system and therefore it is reasonable to assume as trial solution of Eqs.~(\ref{lle}),
 $  \delta a = A_+ \mathrm{e}^{\mathrm{i}\Omega t}+ A_- \mathrm{e}^{-\mathrm{i}\Omega t}$,
and $ \delta q  = X \mathrm{e}^{-\mathrm{i}\Omega t} + \mathrm{c.c.} $. The resulting amplitudes are given by
\begin{equation}
\label{amen}
   A_\pm = \frac{\sqrt{2\kappa_0} s_{\mathrm{p}}}{\kappa_{\mathsmaller{\mathrm{T}}}+\mathrm{i} (\Delta-\Omega)}\left[\delta_{\pm 1,-1}+\mathrm{i}\frac{G^2 \chi_{{\rm eff}}(\mp\Omega)/2}{\kappa_{\mathsmaller{\mathrm{T}}}+\mathrm{i} (\Delta\pm\Omega)}\right],
   \end{equation}
 $ X = \sqrt{\kappa_0} s_{\mathrm{p}} G \chi_{{\rm eff}}(\Omega)/\left[\kappa_{\mathsmaller{\mathrm{T}}}+\mathrm{i} (\Delta-\Omega)\right]$, where $\kappa_{\mathsmaller{\mathrm{T}}} = \kappa_0+\kappa_2$ is the total cavity decay rate, we have introduced the effective optomechanical coupling $ G = -\sqrt{2}\partial_q \omega(q_{\mathrm{s}})\alpha_{\mathrm{s}} $~\cite{Genes2009} given by
\begin{equation}
G = -2\left(\frac{\partial\omega}{\partial z_0}\right)\Theta \sqrt{\frac{\mathcal{P}\kappa_0 }{m \Omega_{\mathrm{m}} \omega_{\mathsmaller{\mathrm{L}}}\left(\kappa_{\mathsmaller{\mathrm{T}}}^{2}+\Delta ^{2}\right) }}, \label{optoc}
\end{equation}
with ${\mathcal P}$ the pump input power, and
\begin{equation}
\chi _{\rm eff}(\omega ) = \Omega_{\mathrm{m}}\Biggl[\tilde{\Omega}_\mathrm{m}^{2}-\omega^{2}-\mathrm{i}\omega \gamma _\mathrm{m}-\frac{G^2\Delta\Omega _\mathrm{m}}{\left(\kappa_{\mathsmaller{\mathrm{T}}} -\mathrm{i}\omega
\right)^{2}+\Delta ^{2}}\Biggr]^{-1},\label{chieffD}
\end{equation}%
is the mechanical susceptibility modified by the optomechanical coupling, with $
\tilde{\Omega}_\mathrm{m}^{2} = \Omega_\mathrm{m}^{2}+h \Omega_\mathrm{m},\quad h = \partial_{q}^2 \omega(q_\mathrm{s})|\alpha_\mathrm{s}|^2$,
the square of the mechanical frequency modified by the second order contribution to the expansion of $\omega(\hat{q})$.

The output field transmitted by the cavity is given by
\begin{equation}\label{transm}
    a_2^{\mathrm{out}}= \sqrt{2\kappa_2}\left(\alpha_{\mathrm{s}}+A_- \mathrm{e}^{-\mathrm{i}\Omega t} +A_+ \mathrm{e}^{\mathrm{i}\Omega t}\right);
\end{equation}
as discussed above, in our setup OMIT manifests itself as a complete reflection of the probe beam by the cavity, even if at resonance. This happens when $A_-=0$, which is realized when the probe is resonant with the cavity and with the blue sideband of the pump, $\Omega\approx \Delta \approx \Omega_{\rm m}\approx \tilde{\Omega}_{\rm m}$, which is analogous to the two-photon resonant condition of usual EIT~\cite{Weis2010,Safavi-Naeini2011,Arimondo1996,Fleischhauer2005,Agarwal2010}. In such a case, in fact, $A_- \propto 1+\mathrm{i}G^2 \chi_{\rm eff}(\Delta)/2\kappa_{\mathsmaller{\mathrm{T}}} \approx 0,$ where the latter condition is realized when the \emph{cooperativity} $C= G^2/2\kappa_{\mathsmaller{\mathrm{T}}}\gamma_{\rm m}$ is sufficiently large, $C \gg 1$, and we are in the resolved sideband regime $\kappa_{\mathsmaller{\mathrm{T}}} \ll \Omega_{\rm m}$, conditions which are both met in our experiment.

In Refs.~\cite{Weis2010,Safavi-Naeini2011,Teufel2011,Massel2012} OMIT is shown by measuring the probe transmission as a function of $\Omega$. Here we show its occurrence in a slightly different way, by measuring the intensity and the phase shift of the beat at frequency $\Omega$ between the transmitted pump and probe fields, $A_{{\rm beat}}$. Using Eq.~(\ref{transm}) and neglecting the field oscillating at $-\Omega$ which is well out of resonance, one gets that the beat amplitude at frequency $\Omega$ of the transmitted field is given by $ A_{{\rm beat}} = 2\kappa_2  \alpha_{\mathrm{s}} A_- $, namely,
\begin{equation}\label{eq:beatamp}
   A_{{\rm beat}} = \frac{4\kappa_2 \kappa_0 |s_{\mathrm{p}}|}{\kappa_{\mathsmaller{\mathrm{T}}}} \sqrt{\frac{{\cal P}}{\hbar \omega_{\mathsmaller{\mathrm{L}}} \left(\kappa_{\mathsmaller{\mathrm{T}}}^2+\Delta^2\right)}}\left[1+ \mathrm{i} \frac{G^2 \chi_{{\rm eff}}(\Delta)}{2\kappa_{\mathsmaller{\mathrm{T}}}}\right],
\end{equation}
where the phase of $A_{\rm beat}$ is referred to the phase of the probe $s_{\mathrm{p}}$, and we have put $\Omega = \Delta$ in Eq.~(\ref{eq:beatamp}) because we have taken the weak probe to be always resonant with the cavity.

The behavior of the measured beat amplitude is shown in Fig.~2, where its phase and modulus are plotted vs the pump-probe detuning $\Omega$, which is kept equal to the cavity-pump detuning $\Delta$. The data refer to an incident pump power ${\mathcal P}\approx3$ mW, and we have independently measured a total cavity rate $\kappa_{\mathsmaller{\mathrm{T}}} \approx 85$ KHz, and an effective mass $m\approx45$ ng. Both plots are in good agreement with the theoretical prediction of Eq.~(\ref{eq:beatamp}) for an effective optomechanical coupling $|G|=9.4 \times 10^{-3} \Omega_{\rm m}$ (full blue line), corresponding to a membrane shifted by $z_0=4 $ nm along the cavity axis with respect to a field node.

\begin{figure}[h]
   \centering
   \includegraphics[width=.45\textwidth]{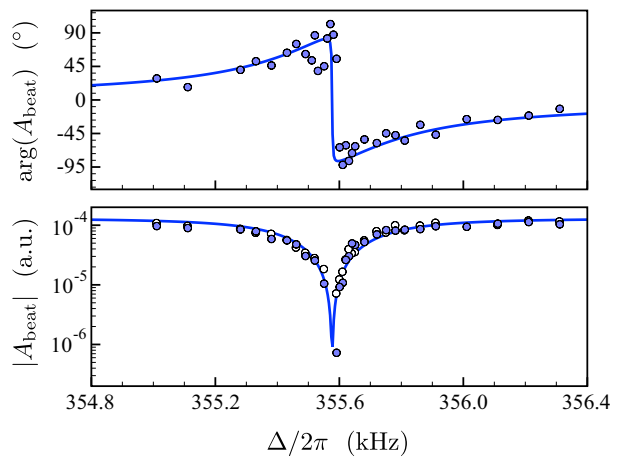}
   \caption{(Color online) Phase shift with respect to the probe (upper panel) and modulus (lower panel) of the beat between the transmitted pump and probe beams vs the pump-probe detuning, which is kept equal to the cavity-pump detuning $\Delta$. The blue full line refers to the theoretical prediction of Eq.~(\ref{eq:beatamp}) with parameters given in the text.}
   \label{fig:omit}
\end{figure}

Fig.~2a shows the phase shift acquired by the probe beam during its transmission through the optomechanical cavity. The derivative of such a phase shift gives the group advance due to causality-preserving superluminal effects which a probe pulse spectrally contained within the transparency window would accumulate in its transmission through the cavity. From the fitting curve of Fig.~2a we infer a maximum signal time advance $\tau^{\mathsmaller{\mathrm{T}}} \approx -108$ ms, which is very close to the theoretical maximum time advance achievable at $\Omega = \Delta = \Omega_{\mathrm{m}}$~\cite{Safavi-Naeini2011}
$\tau^{\mathsmaller{\mathrm{T}}}_{\rm max} =-2C/[\gamma_{\mathrm{m}}(1+C)]$, which is $-109$ ms in our case where $C = 160$. The reflected field is instead delayed, and from the corresponding expression for the maximum time delay  $\tau^{\mathsmaller{\mathrm{R}}}_{\rm max}= 2 \eta C/[\gamma_{\mathrm{m}}(1+C)(1-\eta+C)]$ ($\eta=2\kappa_0/\kappa_{\mathsmaller{\mathrm{T}}}\approx 1$), we can also infer a group delay of the reflected probe field $\tau^{\mathsmaller{\mathrm{R}}}\approx 670$ $\mu$s.

In Fig.~2b the ``transparency'' frequency window in which the probe is \emph{completely reflected} by the interference associated with the optomechanical interaction is evident. The width of the transparency window is related to the effective mechanical damping $\gamma_{\mathrm{m}}^{\rm eff}$, which is approximately given by $\gamma_{\mathrm{m}}^{\rm eff}\approx \gamma_{\mathrm{m}}(1+C)$ around the resonant condition $\Omega_{\rm m}=\Delta=\Omega$ we are considering~\cite{Weis2010,Safavi-Naeini2011} [see also Eq.~(\ref{eq:beatamp})], and therefore increases for increasing cooperativity. This is illustrated in Fig.~3, where the modulus of the beat amplitude vs $\Delta=\Omega$ is plotted for different positions shifts $z_0$ of the membrane from a field node: $z_0=5$ nm (red circles), $z_0=7$ nm (light green up-pointing triangles), $z_0=15$ nm (blue squares), $z_0=21$ nm (orange down-pointing triangles), corresponding to increasing values of the coupling, $|G|/\Omega_{\mathrm{m}} =1.0 \times 10^{-2}$, $1.4 \times 10^{-2}$, $3.1 \times 10^{-2}$, $4.2 \times 10^{-2}$, respectively. The other parameters are the same as those of Fig.~2 except for the mechanical quality factor, which was smaller, $Q=24\,000$, due to the lower quality of the vacuum in the chamber. The data are in very good agreement with the prediction of Eq.~(\ref{eq:beatamp}) (full lines). The inset in Fig.~3 explicitly shows how the EIT bandwidth can be tuned with membrane position $z_0$, at fixed input laser power.

\begin{figure}[h]
   \centering
  \includegraphics[width=.4833\textwidth]{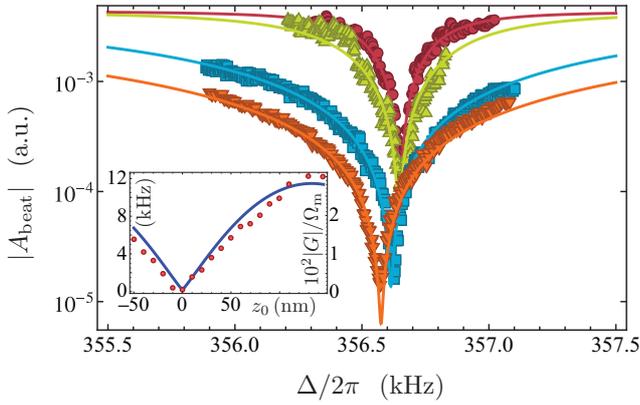}
   \caption{(Color online) Modulus of the beat between the transmitted pump and probe beams vs the pump-probe detuning, which is kept equal to the cavity-pump detuning $\Delta$ for different membrane shifts $z_0$ with respect to a cavity node, as explained in the text. The full lines refer to the theoretical prediction of Eq.~(\protect\ref{eq:beatamp}). The inset shows the width of the EIT window versus the membrane position $z_0$, with respect to a field node, at fixed input laser power.}
   \label{fig:omit2}
\end{figure}

The results show that thermal noise does not have any relevant effect on the EIT window, even if the experiment is carried out at room temperature, at very large mean thermal phonon number $n_{\mathrm{th}} \approx 10^8 \gg C$. There is however an important limitation which occurs in this high temperature limit: the setup can be used to delay and store light pulses carrying only \emph{classical} states but not \emph{quantum} states. In fact only pulses with bandwidth narrower than the EIT window $\approx \gamma_\mathrm{m} C$ can be delayed and stored; at the same time a quantum state is decohered at the thermal decoherence rate $\gamma_\mathrm{m} n_{\mathrm{th}}$, and therefore it can be safely stored only if $n_{\mathrm{th}} < C$.

\emph{Optomechanically induced amplification}. We have then investigated the situation where the pump is resonant with the \emph{blue} sideband of the cavity mode, \emph{i.e.}, when $\Delta=-\Omega_{\rm m}$. In such a case, the probe beam \emph{constructively} interferes with the Stokes sideband of the pump beam which is resonant with the cavity, and may be amplified in transmission within a very narrow frequency window. This is the optomechanical analogue of electromagnetically induced amplification~\cite{Safavi-Naeini2011,Massel2011}, demonstrated in the unresolved sideband limit in~\cite{Verlot2010,McRae2012,Li2012}, and closely related to the electromagnetically induced absorption observed in atomic gases~\cite{Lezama1999}. The latter consists in the decrease of the total power at the output of the medium for increasing pump power, and may occur only when the medium internal losses are larger than the external losses. In our case internal losses are negligible, $\kappa_{\mathsmaller{\mathrm{T}}}=\kappa_0 + \kappa_2 = \kappa_{\rm ext}$, and we may only observe amplification. This can be seen using Eq.~(\ref{eq:beatamp}) for deriving the probe transmission $t_{\rm p}$, which at the blue sideband resonance $\Omega = \Delta \approx -\Omega_{\rm m}$ reads $ t_{\rm p} = \eta'/(1-C)$,
where $\eta'=2\sqrt{\kappa_0 \kappa_2}/\kappa_{\mathsmaller{\mathrm{T}}}$. $t_{\rm p} $ gives the amplifier gain and therefore the probe is amplified in transmission when $C> 1-\eta'$, which is practically always satisfied because in our setup $\kappa_0 \approx \kappa_2 \approx \kappa_{\mathsmaller{\mathrm{T}}}/2$. The amplification bandwidth is narrow and given by the effective mechanical damping in this blue sideband driving condition $\gamma_{\mathrm{m}}^{\rm eff}\approx \gamma_{\rm m}(1-C)$. The amplification of the transmitted probe beam is well visible in Fig.~4, where the modulus of the beam amplitude is plotted vs $\Delta$, but now around the condition $\Delta=-\Omega_{\rm m}$. The full line corresponds to the prediction of Eq.~(\ref{eq:beatamp}). Fig.~4 refers to an input power ${\cal P}\approx50$ $\mu$W, a membrane shifted by $z_0\approx5$ nm from a node, corresponding to a coupling $|G|/\Omega_{\mathrm{m}} \approx 10^{-3}$, and a quality factor $Q\approx24\,000$, yielding in this case $C \approx 0.32$.

\maybeflushthispage
\begin{figure}[h]
   \centering
   \includegraphics[width=.45\textwidth]{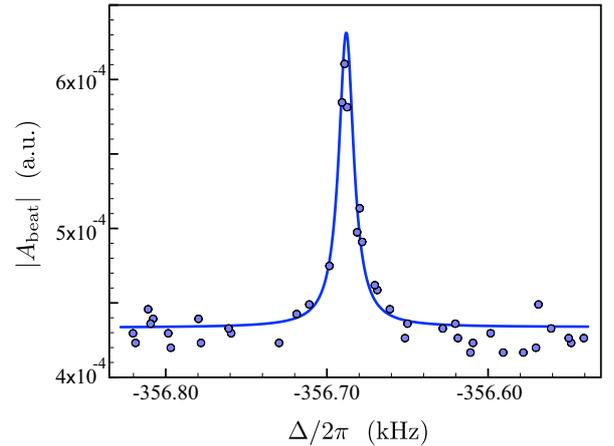}
   \caption{(Color online) Modulus of the beat between the transmitted pump and probe beams vs the pump-probe detuning, in the case of optomechanically induced amplification. The full line refers to the theoretical prediction of Eq.~(\ref{eq:beatamp}). See text for parameters.}
   \label{fig:omia}
\end{figure}

The system is stable as long as $\gamma_{\mathrm{m}}^{\rm eff} >0$, \emph{i.e.}, only if $C<1$. In this regime the system is the optomechanical analogue of a parametric oscillator below threshold. The system has been studied even at larger cooperativity and the nonlinear amplification process controlled by membrane position along the optical axis has been observed. At large cooperativity few mW of pump power have been transferred to the cavity resonance. This process, where the mechanical resonator starts to oscillate with a nonzero amplitude, has been theoretically discussed in~\cite{Marquardt2006,Rodrigues2010} and experimentally demonstrated in~\cite{Kippenberg2005,Metzger2008}.

\emph{Acknowledgments}. This work has been supported by the European Commission (ITN-Marie Curie project cQOM).

\bibliography{optomechanics-eit}

\end{document}